\documentclass[12pt,preprint]{aastex}

\shorttitle{Distance of Cr 121 }
\shortauthors{Kaltcheva and Makarov}
\begin{document}

\title{The Structure and the Distance of Collinder 121  from $Hipparcos$ and Photometry:\\ Resolving the Discrepancy}
\author{Nadia Kaltcheva}
\affil{Department of Physics and Astronomy, University of Wisconsin Oshkosh, 800 Algoma Blvd., Oshkosh, WI 54901}
\email {kaltchev@uwosh.edu}
\and

\author{Valeri Makarov}
\affil{Michelson Science Center, California Institute of Technology, 770 S. Wilson Ave., MS 100-22,   Pasadena, CA 91125}
\email {vvm@caltech.edu}
\begin{abstract}
We present further arguments that the $Hipparcos$ parallaxes for some
of the clusters and associations represented in the $Hipparcos$
catalog should be used with caution in the study of the Galactic
structure. It has been already shown that the discrepancy between the
$Hipparcos$ and ground based parallaxes for several clusters including
the Pleiades, Coma Ber and NGC 6231 can be resolved by recomputing the
$Hipparcos$ astrometric solutions with an improved algorithm
diminishing correlated errors in the attitude parameters. Here we
present new parallaxes obtained with this algorithm for another group
of stars with discrepant data - the galactic cluster Cr 121. The
original $Hipparcos$ parallaxes led de Zeeuw et al. to conclude that
Cr 121 and the surrounding association of OB stars form a relatively
compact and coherent moving group at a distance of $\simeq 550$ -- 600
pc. Our corrected parallaxes reveal a different spatial distribution
of young stellar populace in this area. Both the cluster Cr 121 and
the extended OB association are considerably more distant ($750$ --
$1000$ pc), and the latter has a large depth probably extending beyond
1 kpc. Therefore, not only are the recalculated parallaxes in complete
agreement with the photometric $uvby\beta$ parallaxes, but the
structure of the field they reveal is no longer in discrepancy with
that found by the photometric method.

\end{abstract}


              
\def\Stromgren{Str{\"o}mgren }
\def\o{\"o }
\def\V0{$V_0$ }
\def\V{$V$}
\def\eby{$E(b-y)$ }
\def\ebv{$E(B-V)$ }
\def\Mv{$M_V$ }
\def\Av{$A_V$ }
\def\b-y{$b-y$ }
\def\c1{$c_1$ }
\def\m1{$m_1$ }
\def\BrM1{$[m_{1}]$ }
\def\BrC1{$[c_{1}]$ }
\def\d{$ ^{\rm o}$ }
\def\'{$ ^{\rm '}$ }
\def\C0{$c_0$ }
\def\M0{$m_0$ }
\def\by0{$(b-y)_0$ }

\section{Introduction}
 
Obtaining reliable knowledge about the structure and distance of
nearby OB associations plays a critical role in the overall study of
the Milky Way morphology near the Sun. Unlike the external galaxies
where the star-forming fields are generally evident from direct
imaging, the study of the spiral structure of our own Galaxy is
largely grounded in distance determinations of young stellar
tracers. At present, sufficiently accurate astrometric data
(parallaxes and proper motions) are available for few star-forming
regions within $\simeq 500$ pc. More comprehensive and representative
studies of the local history and dynamics of star formation have to
rely on the photometric method of distance determination and stellar
evolution theory.

The completion of the $Hipparcos$ catalog (ESA 1997) offered a
possibility for a major improvement of the membership of young moving
groups near the Sun and refining the distance scale
to nearby open clusters and OB associations. However, the mean
$Hipparcos$ parallaxes for some galactic clusters are in disagreement
with ground-based determinations by various methods. Statistically
significant discrepancies between the $Hipparcos$ trigonometric and
traditional photometric, spectroscopic and interferometric results
have been reported in the literature for selected small-scale fields,
most notably for the Pleiades open cluster (Pinsonneault et al. 1998,
Soderblom et al. 1998, Narayanan \& Gould 1999, Stello \& Nissen 2001,
Makarov 2002, Pan et al. 2004, Percival et al. 2005, Soderblom et al. 2005). Platais et
al. (2007) found a similar offset in the $Hipparcos$ mean parallax for
the young open cluster IC 2391. A discrepancy was reported by
Kaltcheva et al. (2005) for the open cluster IC 2602 as well. The
cause for these inconsistencies is most likely due to a faulty data
reduction algorithm used in $Hipparcos$, which allowed highly
correlated errors of along-scan attitude parameters to propagate into
the fitted astrometric parameters. An alternative data reduction
approach has been suggested and successfully tested by Makarov (2002,
2003).

The region of Cr 121 is another example of this discrepancy. Since the
discovery of a compact group at $\ell,b$= $(234.98\degr, -10.21\degr)$
by Collinder (1931), both the cluster and the larger $10\degr$ x
$10\degr$ field have been extensively studied by UBV and $uvby\beta$
photometry. This area includes one of the 12 OB associations within 1
kpc from the Sun with fairly detailed kinematical information and
membership determined from $Hipparcos$. The $Hipparcos$ proper motions
reveal a moving group of 103 stars between $\ell=227\degr$ and
$\ell=245\degr$, identifying the compact cluster Cr 121 with an unbound
extended OB association at a distance of 592$\pm$28 pc, similar to Sco
OB2 (de Zeeuw et al. 1999). Robichon et al. (1999) selected 13
$Hipparcos$ members of Cr 121 and found a mean $Hipparcos$ parallax of
1.80$\pm$0.24 mas ($556\pm 74$ pc). In contrast to these results from
$Hipparcos$, the latest $uvby\beta$ photometric study (Kaltcheva 2000)
concluded that a compact stellar group apparently identical to the
genuine cluster (Cr 121) is situated at 1085$\pm$41 pc and the closest
members of the loose association are found at an average distance of 660-730
pc, in agreement with most of the previous photometric investigations.
Since the $uvby\beta$ photometry is arguably the best photometric
system in use to provide accurate photometric distances, the origin of
the discrepancy was suggested to be in the $Hipparcos$ parallaxes for
the Cr 121 members.  Burningham et al. (2003) studied the low-mass
pre-main sequence stars toward Cr 121 and also came to conclusions
consistent with the photometric distance determinations.

In this letter, we consider a sample of probable members of the
extended association around Cr 121 selected by de Zeeuw et al. (1999) for which
accurate $uvby\beta$ photometry is available. The astrometric
parameters of these stars are recomputed from the $Hipparcos$
Intermediate Astrometry Data by the method proposed by Makarov
(2002). The recomputed parallaxes allow us to resolve the controversy
about the distance and dimensions of the OB association in this field.

\section{Results and Discussion}
Our sample contains all 44 early-type stars with $Hipparcos$ parallaxes listed by
de Zeeuw et al. (1999) as probable members of the Cr 121 moving group for which
$uvby\beta$ photometry is available. Homogeneous photometric
$uvby\beta$ distances are calculated for 43 of them (Kaltcheva 2000).
Table 1 presents the sample, where the $Hipparcos$ identification
numbers are given in the first column, followed by the $Hipparcos$
parallaxes and their errors, recalculated parallaxes and their errors,
photometric $uvby\beta$ distances and MK spectral classification. The
stars are formally divided into field stars (or possible association members),
spread over a $10\degr\times 10\degr$ area around the center of Cr 121
and 6 photometrically selected members of the dense cluster Cr 121
(Kaltcheva 2000 and references therein). As follows from the data in Table 1 
there is a statistically significant difference between the mean $Hipparcos$ parallax of 
$1.87\pm0.15$ mas and the mean recomputed parallax of $1.29\pm0.15$ mas. The errors provided here are the  formal standard deviation of the mean computed  from the formal errors of parallaxes.

Fig.~1 shows the original $Hipparcos$ parallaxes (left plot) and our
recomputed parallaxes (right plot) versus the photometric parallaxes
for the sample of 43 stars in Table ~1. The $Hipparcos$ parallaxes are
on average larger than the photometric values by 0.52$\pm$0.107 mas, where the quoted error is the sample standard error of the mean. This is a statistically significant difference of the same
order as those found for the Pleiades and a few other Galactic
clusters. On the other hand, the agreement is excellent between the
mean photometric parallax and the mean corrected parallax ($0.063\pm
0.158$ mas).  This supports our main conclusion that the $Hipparcos$
parallaxes are systematically overestimated in this area of the
sky. But Fig.~1 also reveals another strange property of the original
parallaxes. While the recomputed parallaxes are scattered fairly
symmetrically around the line of unit slope in the right plot and
their dispersion is in good agreement with the measurement errors, the
original parallaxes are grouped tightly around the mean ($1.87$ mas)
with a standard deviation of only $0.61$ mas. This value is much too
small for the estimated formal errors (mean $0.93$ mas, {\it rms}
$0.96$ mas).  We attribute this result to a strong selection effect in
the method employed by de Zeeuw et al. (1999), which preferentially accepted
stars with large measured parallaxes, i.e., mostly stars with positive
errors "observed minus true". In combination with the correlated error
of the mean parallax, this selection bias gives rise to doubt about
the completeness and reliability of the present membership list.

The significant dispersion of both photometric and recomputed
parallaxes also implies a complex morphology of this moving group
having a considerable depth, as opposed to the previous conjecture of
an association compressed in the radial dimension, similar to the
nearby Sco OB2 association, as concluded by de Zeeuw et al.
(1999).  The group also appears to be located at a larger distance of  $\simeq
740$ pc, rather than at $\simeq 550$ pc as follows from the mean $Hipparcos$
parallax of the sample in Table 1.  Based on a larger photometric sample
it has already been pointed out that the loose nearby structure
defined by de Zeeuw et al. (1999) to be located at 592$\pm28$ pc
photometrically appears to be more distant by about 100 pc (Kaltcheva
2000).  The parallaxes recalculated here support the photometric findings.

Our result implies that the problem of inaccurate mean
parallaxes in $Hipparcos$ affects more regions, and of larger angular
area, than just a few small patches occupied by dense open
clusters. This is not an irreversible situation, because the method of
astrometric solution of the available $Hipparcos$ data used in this
paper proves once again successful in correcting this error, despite
its limitations.  A more systematic and thorough comparison of
$Hipparcos$ data with distances from precision multi-band photometry
will probably reveal more problematic areas. It is not clear at
present how widely spread the parallax error is, and whether a global
astrometric solution will have a significant impact on the present
knowledge of distances and morphology for many of the OB associations
represented in the catalog, but it is evident that the
$Hipparcos$-based census of some of the moving groups near the Sun
should be critically reconsidered.

\clearpage
\begin{table}[here!]
\caption{The sample: identifications from the $Hipparcos$ catalog, followed by  the $Hipparcos$ parallaxes $\pi$ and their formal errors $\sigma_\pi$, the recalculated parallaxes $\pi_r$ and their formal errors $\sigma_{\pi_r}$,   the photometric $uvby\beta$ distances r  and the MK classification.}
\scriptsize
\vspace{0.1in}
\begin{tabular}{lrrrrrl}
\hline
\hline
HIP   & $\pi$   &   $\sigma_\pi$   &  $\pi_r$   &   $\sigma_{\pi_r}$    &    r     &       MK    \\ 
   & (mas)  &   (mas)  &   (mas)  &  (mas)   &     (pc)    &       \\ 
       &       & 	  & 	   & 	     	  & 	   & 	            \\
Field stars       &       & 	  & 	   	& 	  & 	   & 	            \\
31436  & 1.14  &   0.90   &  0.67  &   0.94   &    1812  &   B2/B3V     \\
31901  & 2.05  &   1.06   &  0.95  &   1.07   &   1050  &   B5         \\ 
32084  & 2.63  &   1.18   &  -1.02 &   1.09   &    664   &   B3V        \\
32101  & 1.24  &   1.07   &  -0.21 &   1.11   &   938   &   B9.5III    \\
32591  & 1.42  &   0.97   &  0.14  &   0.98   &   605   &   B8V        \\
33007  & 1.16  &   0.86   &  1.17  &   0.92   &   475   &  B4V          \\
33092  & 2.02  &   0.70   &  2.37  &   0.78   &    518   &   B1Ib        \\
33165  & 1.74  &   0.76   &  1.33  &   0.85   &    -     &   WN...       \\
33260  & 1.19  &   1.10   &  0.98  &   1.11   &  930   &   B9Ib/II     \\
33294  & 1.43  &   0.69   &  0.82  &   0.77   &   681   &   B2III/IV    \\
33316  & 1.51  &   0.64   &  -0.32 &   0.73   &    632   &   B2/B3III    \\
33447  & 2.78  &   0.70   &  1.23  &   0.77   &   766   &   B2III/IV    \\
33523  & 1.70  &   1.23   &  -0.41 &   1.25   &   1697  &   B2V         \\
33532  & 2.24  &   0.73   &  1.04  &   0.85   &    539   &   B2.5III     \\
33611  & 2.05  &   0.70   &  1.40  &   0.76   &    722   &   B2V         \\
33621  & 1.70  &   0.93   &  0.29  &   0.98   &    764   &   B8II/III    \\
33666  & 2.33  &   0.68   &  0.90  &   0.76   &   740   &   B2III       \\
33673  & 1.68  &   0.72   &  0.56  &   0.78   &    923   &   B4Vn        \\
33721  & 2.46  &   0.74   &  1.43  &   0.81   &    706   &   B3Vnn       \\
33769  & 1.26  &   0.80   &  0.25  &   0.85   &    1077  &   B2/B3V      \\
33770  & 2.05  &   0.97   &  1.31  &   1.28   &    630   &   B2IV        \\
33804  & 3.17  &   0.59   &  3.29  &   0.66   &  365   &   B2/B3III/IV \\ 
33814  & 2.31  &   0.93   &  2.44  &   0.97   &    887   &   B3V          \\
33846  & 1.41  &   0.74   &  2.04  &   0.80   &   647   &   B3V          \\
33865  & 1.75  &   1.18   &  -0.14 &   1.35   &  648   &   B3IV         \\
33888  & 1.35  &   1.13   &  1.38  &   1.14   &   793   &   B9V+...      \\
34041  & 1.79  &   0.66   &  1.48  &   0.72   &   521   &   B2/B3V       \\
34067  & 1.66  &   0.80   &  2.27  &   0.84   &    853   &   B3III        \\
34074  & 1.10  &   1.10   &  1.82  &   1.13   &    1597  &   B7/B8III     \\
34153  & 2.55  &   1.06   &  1.92  &   1.09   &    535   &   B8V          \\
34167  & 1.44  &   0.91   &  1.45  &   0.94   &    958   &   B2IV         \\
34219  & 1.95  &   1.67   &  2.94  &   1.38   &   665   &   B6III        \\
34227  & 1.04  &   0.94   &  0.90  &   0.97   &   757   &   B3V:n        \\
34281  & 1.28  &   1.03   &  1.16  &   1.04   &    842   &   B5V          \\
34331  & 2.23  &   0.65   &  1.11  &   0.71   &    534   &   B2IV-V       \\
34579  & 1.78  &   0.60   &  1.64  &   0.68   &   368   &   B2V          \\
34940  & 2.07  &   1.24   &  3.18  &   0.98   &   676   &   B2IV         \\
35026  & 1.44  &   0.78   &  2.34  &   0.83   &   1435  &   B2IV/V       \\
       &       & 	  & 	   & 	      &	   & 	            \\
Cr 121 &       & 	  & 	   & 	      & 		   & 	            \\
32823  & 1.92  &   1.23   &  2.41  &   1.26   &    944   &   B5V          \\
32911  & 3.49  &   1.02   &  2.75  &   1.05   &   1012  &   B8IV/V       \\
33062  & 1.22  &   0.96   &  0.81  &   1.02   &   947   &   B2II/III     \\
33070  & 2.30  &   1.13   &  0.97  &   1.17   &    1291  &   B3II/III     \\
33208  & 1.77  &   1.14   &  1.16  &   1.16   &   981   &   B3V          \\
33211  & 3.46  &   1.08   &  2.51  &   1.11   &    1131  &   B3V          \\										     
\hline
\hline
\end{tabular}
\end{table}

\clearpage
\begin{figure}[here!]
\epsscale{1}
\plottwo{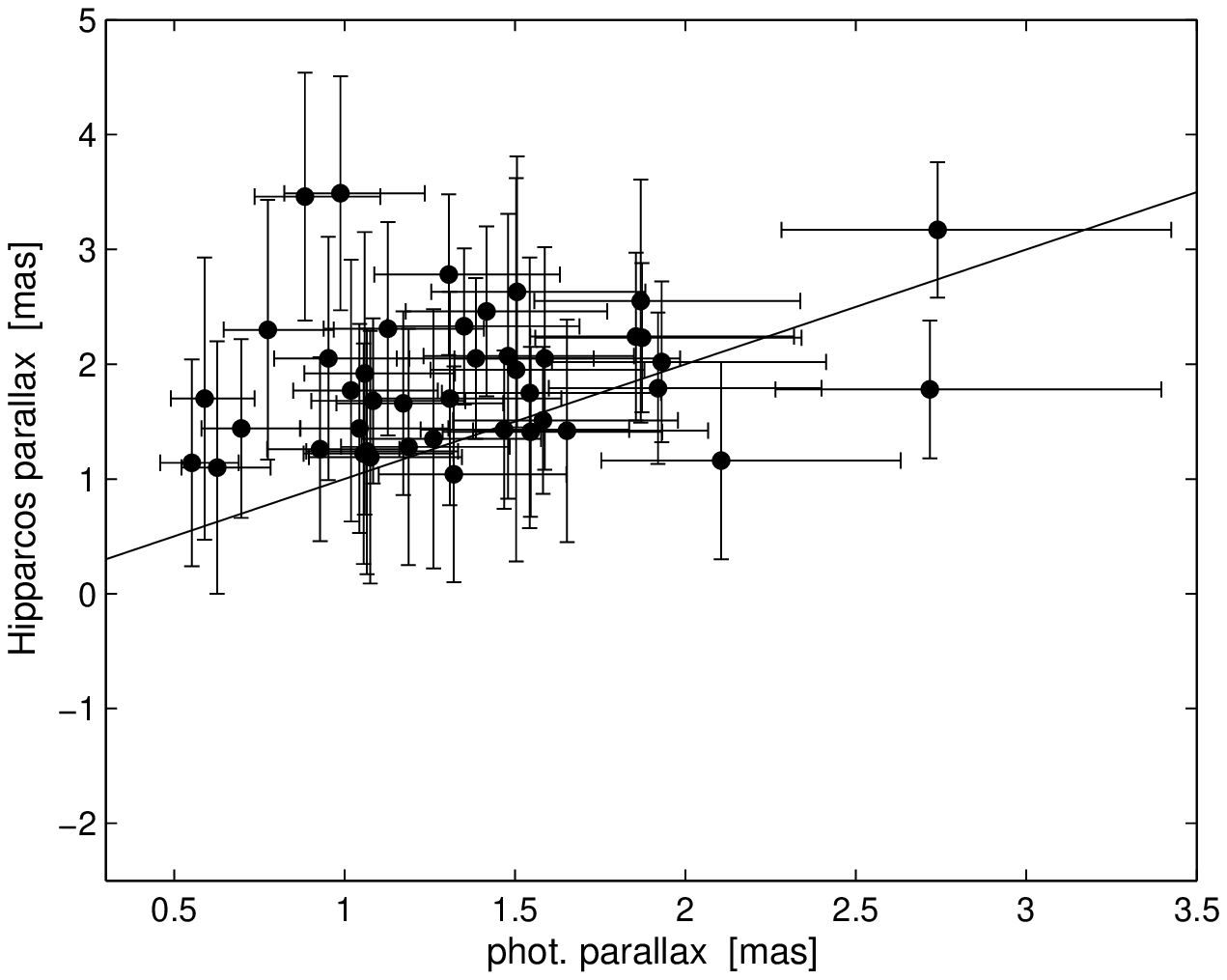}{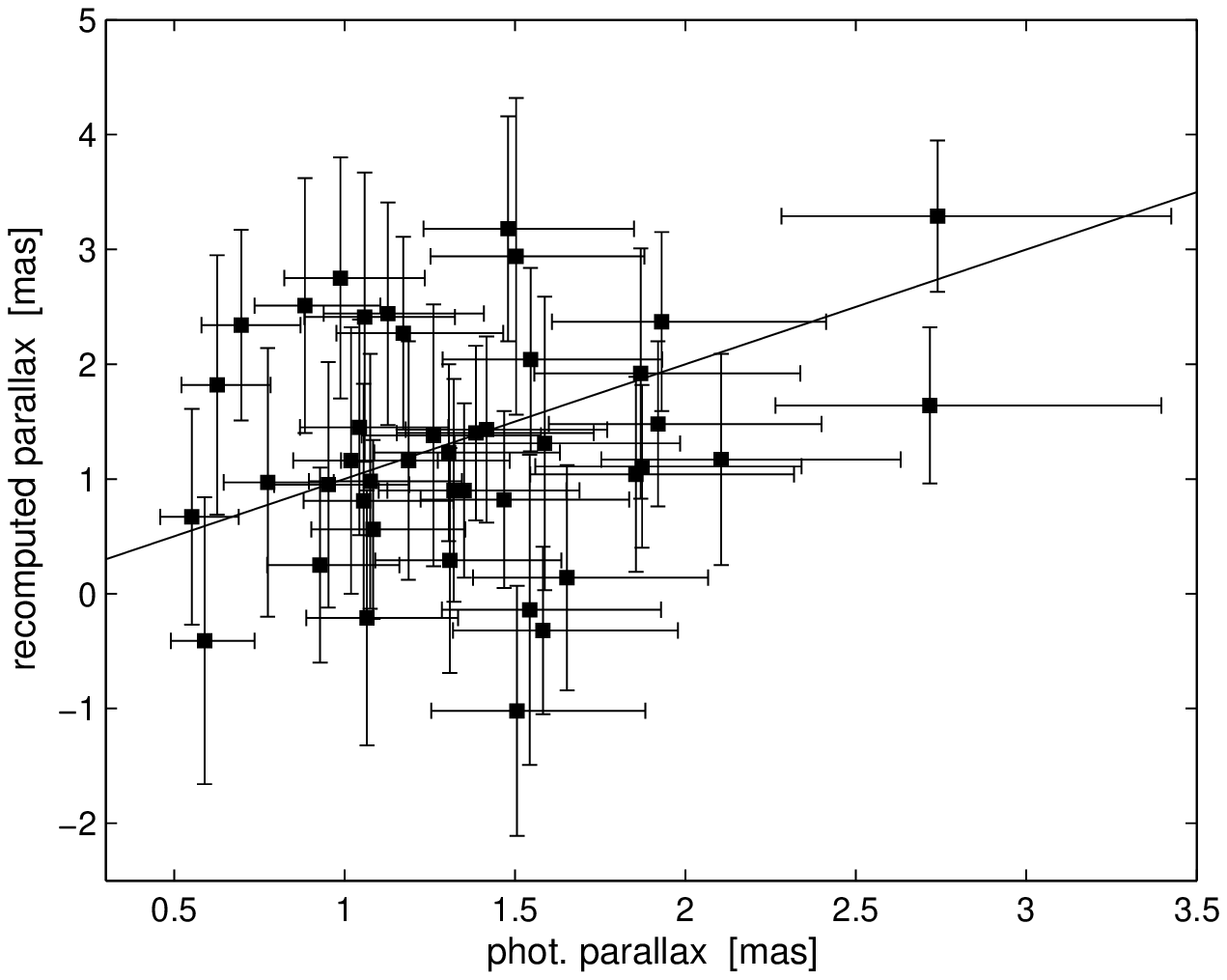}
\caption{Differences  between the $uvby\beta$ photometric parallaxes and $Hipparcos$ parallaxes (left plot), and the parallaxes recomputed in this paper (right plot) for stars in the area of Cr 121. The error bars of the photometric parallaxes correspond to the maximum estimated error in the photometric distances of 20 \%.  }
\end{figure}
\begin{acknowledgements}

\end{acknowledgements}
We are grateful to the referee Dr. R. Hanson for a number of valuable comments. A UW Oshkosh Vander Putten Award is acknowledged. The research described in this paper was in part carried out at the Jet Propulsion Laboratory,
California Institute of Technology, under a contract with the National Aeronautics and Space
Administration.

\begin{thebibliography}{}
\bibitem[1978]{BDe}Burningham, B., Naylor, T., Jeffries, R. D., Devey, C. R., 2003, MNRAS, 346, 1143
\bibitem[1978]{Dee}Collinder, P., 1931, Ann. Obs. Lund, 2, 1
\bibitem[de Zeeuw et al.(1999)]{De}de Zeeuw, P.T., Hoogerwerf, R., de Bruijne, J.H.J., Brown, A.G.A., \& Blaauw, A., 1999, \aj, 117, 354
\bibitem[1997]{ES}ESA, 1997, The Hipparcos Catalogue, ESA SP-1200, Vol. 1-17
\bibitem[1978]{Si}Kaltcheva, N.T., 2000, MNRAS, 318, 1023
\bibitem[1978]{Se}Kaltcheva, N.T., Jaeger, S., Kaba Bah, M., \& Briley, M.M. 2005, AN, 326 ,738
\bibitem[1978]{Ss}Makarov, V.V.,  2002, AJ, 124, 3299
\bibitem[1978]{Sg}Makarov, V.V.,  2003, AJ,126, 2408
\bibitem[Pan et al. (2004)]{pan} Pan, X., Shao, M., \& Kulkarni, S.R. 2004, Nature, 427, 326
\bibitem[Percival et al. (2005)]{per} Percival, S.M., Salaris, M., \& Groenewegen, M.A.T. 2005, \aap, 429, 887
\bibitem[Pinsonneault et al.(1998)]{pin} Pinsonneault, M.H., Stauffer, J., Soderblom, D.R., King, J.R., \& Hanson, R.B. 1998, \apj, 504, 170
\bibitem[Platais et al.(2007)]{pla07} Platais, I., Melo, C., Mermilliod, J.-C., Kozhurina-Platais, V., Fulbright,J.P., Mendez, R.A., Altmann, M., \& Sperauskas, J.,  2007, \aap, 461, 509
\bibitem[1978]{S}Robichon, N., Arenou, F., Mermilliod, J.C., \& Turon, C., 1999, A\&A, 345, 471
\bibitem[Soderblom et al.(1998)]{sod} Soderblom, D.R., King, J.R., Hanson, R. B., Jones, B.F., Fisher, D., Stauffer, J.R., \& Pinsonneault, M.,  1998, \apj, 504, 192
\bibitem[Soderblom et al.(2005)]{sod5} Soderblom, D.R., Nelan, E., Benedict, G. F., McArthur, B., Ramirez, I., Spiesman, W., \& Jones, B. F., 2005, \aj, 129, 1616
\bibitem[Stello \& Nissen(2001)]{ste} Stello, D., \& Nissen, P.E. 2001, \aap, 374, 105
\end{thebibliography}
\end{document}